\title{Asset Allocation via Machine Learning and Applications to Equity Portfolio Management}
\documentclass[12pt]{article}
\usepackage[utf8]{inputenc}
\usepackage{amssymb}
\usepackage{amsmath}
\usepackage{mathrsfs}
\usepackage{hyperref}
\usepackage{graphicx}
\usepackage{url}
\usepackage{ragged2e}
\usepackage[utf8]{inputenc}
\usepackage[english]{babel}
\usepackage{parskip}
\usepackage{indentfirst}
\usepackage{siunitx}
\usepackage{dcolumn,booktabs}
\newcolumntype{d}[1]{D{.}{.}{#1}}
\newcommand\mc[1]{\multicolumn{1}{c}{#1}}
\usepackage{geometry}
\usepackage[figurename=Exhibit., tablename=Exhibit.]{caption}
\usepackage{times}
\usepackage{titlesec}
\usepackage{lipsum}
\usepackage{babel}
\newtheorem{theorem}{Theorem}

\newtheorem{assumption}{Assumption}

\makeatletter

\let\c@table\c@figure
\makeatother

\author{
Qing       Yang  \thanks{School of Economics, Fudan University, qyang@fudan.edu.cn. Qing Yang is a Professor of Finance at the School of Economics of Fudan University, Shanghai, China.}
\and
Zhenning   Hong  \thanks{Dongxing Securities, Co., Ltd., Asset Management Division, hongzn@hotmail.com. Zhenning Hong is a Director and the Head of Quantitative Investment at the Asset Management Division of Dongxing Securities, Co., Ltd., Shanghai, China.}
\and
Ruyan      Tian  \thanks{School of Economics, Fudan University, 20110680039@fudan.edu.cn. Ruyan Tian is a Ph.D. student of first year at the School of Economics of Fudan University.}
\and
Tingting   Ye    \thanks{Maine Business School, University of Maine, tingting.ye@maine.edu. Tingting Ye is an Assistant of Professor of Accounting at University of Maine, Maine Business School.}
\and
Liangliang Zhang \thanks{Independent, liangliangzhang81@qq.com. Liangliang Zhang will soon join the team of quantitative investment at the asset management division of Dongxing Securities. Liangliang Zhang is the corresponding author of this article.}
}

\begin{document}
\newpage
\maketitle

\newpage
\thispagestyle{plain}
\begin{center}
\LARGE
Asset Allocation via Machine Learning and Applications to Equity Portfolio Management \\
\vspace{0.4cm}
\large
\date{\today}
\vspace{0.4cm}
\normalsize
\begin{abstract}
In this paper, we document a novel machine learning based bottom-up approach for static and dynamic portfolio optimization on, potentially, a large number of assets. The methodology applies to general constrained optimization problems and overcomes many major difficulties arising in current optimization schemes. Taking mean-variance optimization as an example, we no longer need to compute the covariance matrix and its inverse, therefore the method is immune from the estimation error on this quantity. Moreover, no explicit calls of optimization routines are needed. Applications to equity portfolio management in U.S. and China equity markets are studied and we document significant excess returns to the selected benchmarks.
\vspace{0.5cm}
\begin{itemize}
	\item This paper proposes a fast and convergent numerical framework, which is universal and applies to arbitrary constrained optimization problems with unique solutions without calling explicitly any optimization routines, unlike the current problem-specific deep learning-based methods in the literature. The method enjoys global convergence and will not be trapped in local optima;
	\item Our methodology involves no estimation of cross higher order moments of the asset span by its construction. This is crucial for the methodology to overcome the curse of dimensionality when higher order moments are involved, and the number of assets is very large;
	\item We provide empirical studies of portfolio optimization on hundreds and thousands of stocks in U.S. and China equity markets with exotic objective functions and portfolio constraints and document the performance results.
\end{itemize}
\end{abstract}
\end{center}
\vspace{0.6cm}

\newpage
\section{Introduction}
\label{sec:intro}
\subsection{The Outline of This Paper}
In this paper, we propose a novel Monte-Carlo simulation and machine learning-based (unsupervised and supervised learning) static and dynamic portfolio optimization framework. This framework supports arbitrary objective functions and constraints, which can be either linear or nonlinear. Moreover, the number of assets being considered can be very large. Our methodology is fast, accurate and convergent to the global optimum under minimum assumptions, extending the original method introduced in \cite{Zhang2020a}.

The framework consists of an input preparation module, a hierarchical clustering-based asset space decomposition module, a simulation and portfolio weights selection module and a profit and loss evaluation module. The input preparation module applies supervised-learning approach to estimate the inputs that feed into the portfolio optimization module. The hierarchical or regular clustering-based asset space decomposition module tries to partition the entire asset universe, which often includes a large number of individual assets, based on a predetermined set of risk factors. This results in similar factor values for the assets in each cluster. The third module, which is the simulation and portfolio weights selection module, generates uniformly distributed portfolio weights in the constraint region and selects the globally optimal one corresponding to a certain objective utility function with nonlinear constraints. The method in this module is fast and convergent under minimal assumptions. Most importantly, this module does not require the computation of joint higher order moments\footnote{For example, covariance matrix or the tensor of third order moments.} of a random vector (e.g., the rate of returns), which is computationally expensive especially in high dimensions. The last module evaluates the profit and loss of the selected portfolio and performs backtesting. We empirically test our methodology with numerical examples covering various objective functions and constraints, in U.S. and China stock markets.

\subsection{Literature Review}
Portfolio optimization is an important topic in financial economics that attracts both academic researchers and financial practitioners. Starting from the seminal work of modern portfolio theory in \cite{M1952}, where it was suggested that correlations between different assets should be included as inputs to portfolio optimization practice, in addition to volatility and expected returns, the theory of portfolio selection has become rigorous science rather than art. The modern portfolio theory later inspired the famous CAPM model first proposed in \cite{Treynor1962} and later extended in \cite{Merton1973} to a dynamic setting. Through years, we have witnessed an explosion of the number of results on portfolio optimization as the advancement of theoretical and empirical research, and the increase of computational power. Refinements on modern portfolio theory have been studied in \cite{BDDGL2008} in a context of sparse portfolios that stabilizing the output of mean-variance model. The carrier portfolio strategy, described in \cite{Kusiak2013}, is another approach which can deliver sparse and stable portfolios. It relies on a simple, mathematical linear program that directly considers and treats each return observation as individual data, with no assumptions on joint return distributions. Other examples, in a single period setting, include the Black and Litterman approach that was extended by \cite{BL1992}, in which a Bayesian type analysis was used to incorporate investors’ views that correct the equilibrium CAPM expected asset returns. Because the first order moments of asset returns are notoriously hard to estimate, simple return-agnostic strategies were created to account for this phenomenon. For example, the minimum variance portfolio and equal weight (1-over-N) portfolio are studied in \cite{DPUV2013} and \cite{ALS2012}. In addition to the first and second moments, there have been strategies based on higher order moments, namely, skewness and kurtosis, appearing in the work of \cite{HLLM2010} and  many others. The computation of higher order co-moments may rely on factor models, such as the single-factor method considered in \cite{MZ2010}, or forward-looking information obtained from option prices as documented in \cite{CJC2012}. But whichever approach we choose, the methods suffer greatly from the curse of dimensionality as the dimension of asset span increases. Since the proposal of \cite{Ang2013} on factor investing, factor-based strategies start to emerge. References can be found in \cite{KSS2014} and \cite{RW2012}, among others. Additional return agnostic, or risk-based strategies, such as maximum diversification strategy, risk-parity strategies, and their variations have been proposed. Research results can be found in \cite{ABG2014} and references therein. Another strand of literature focuses on dynamic portfolio allocation in a randomly varying market environment and considers multi-period optimization problems. The pioneering work attributes to \cite{Merton1971} with numerous later refinements, see \cite{CK1992} and \cite{SS1999} for example. The characteristics of this type of problems are that they often involve dynamic programming and in a continuous time setting, a PDE system often needs to be solved.

Theoretically being sound, the Markowitz’s mean-variance optimization method, still popular in financial industry, has many obvious drawbacks, which are well-documented in the literature, e.g., \cite{Homescu2014} and \cite{PR2019}. There are three major difficulties with respect to the mean-variance approach. The first is that the method output is extremely sensitive to the model input, which is often difficult to estimate accurately. The second is that the analytical solution of the quadratic programming problem involved requires the computation of a large covariance matrix and its inverse, when the number of assets is large. Third, the computational burden increases sharply if multiple linear or nonlinear constraints are added. To address the input estimation problem, many deep-learning based methods have been proposed recently, such as \cite{YLW2018}, \cite{GKX2020}, \cite{BB2020} and references therein. To address the second and third questions, \cite{PR2019} reviewed machine learning optimization approaches to solve the constrained quadratic programming problem in high dimensions. In \cite{Homescu2014}, general formulations of static portfolio optimization are outlined, taking into consideration the reward, risk and various constraints on optimal portfolios. Compared to the reference in the literature, our method enjoys all the advantages while being theoretically simple and easy to implement in practice.

\subsection{Our Contributions}
The contributions of this paper are five-folds. First, it proposes a fast and convergent numerical framework, which is universal and applies to arbitrary constrained optimization problems with unique solutions without calling explicitly any optimization routines, unlike the current problem-specific deep learning-based methods in the literature. Second, our methodology involves no estimation of cross higher order moments of the asset span by its construction. This is crucial for the methodology to overcome the curse of dimensionality when higher order moments are involved, and the number of assets is large. Third, the paper proposes to use the (hierarchical) clustering method to reduce the dimension of the optimization problem, when there are many assets in the portfolio. Fourth, the paper advocates using deep learning techniques to estimate the model input. Fifth, we provide empirical studies on portfolio choice among a large number of stocks in both U.S. and China equity markets and provide performance analysis.

\subsection{The Organization of This Paper}
The paper is organized as following. Section \ref{sec:m} describes the methodology and the optimization framework. Section \ref{sec:es} performs numerical experiments and Section \ref{sec:c} concludes.

\section{The Methodology}
\label{sec:m}
In this section, we document and present first three of the aforementioned $4$ modules. We first prepare the inputs for the optimization process, which are often the first and higher order moments of asset return vectors. Second, we decompose the asset span, which usually consists of thousands of assets, with the help of a set of risk factors, into small subsets in a hierarchical manner, perform optimization on each sub-level and obtain the final optimal weight on every asset based on the intermediate weights on each cluster. Third, most importantly, we perform portfolio optimization at each clustering level in the hierarchy based on Monte-Carlo simulation. The last step is to compute the results of the backtesting and present the transaction cost at each point in time. Theoretical convergence results are provided in this section and we will illustrate how to perform portfolio choice numerically for both static and dynamic problems.

\subsection{Input Preparation}
\label{sec:ip}
To compute the $\alpha$th order conditional moments $\mathbb{E}_t \left[ R_{t+h}^\alpha \right]$, where $R = \left(R^1, \cdots, R^n\right)$ denotes the vector of asset returns and $\alpha = \left(\alpha_1, \cdots, \alpha_n\right)$ represents the standard multi-index notation, i.e., $R_{t+h}^\alpha = \prod_{i = 1}^n \left[ R_{t+h}^i \right]^{\alpha_i}$ and $|\alpha| = \sum_{i = 1}^n \alpha_i$, we first look at the general semi-martingale decomposition below, written in matrix and vector notations
\begin{align}
\label{eq:semimartdecomp}
R_{t+h} &= \mathbb{E}_t\left[R_{t+h}\right] + \left( R_{t+h}-\mathbb{E}_t\left[R_{t+h}\right] \right) \\
        &= \mu_t + \sigma_t U_{t,t+h}.
\end{align}
Here the random source term $U_{t,t+h}$ satisfies $\mathbb{E}_t\left[U_{t,t+h}\right] \equiv 0$ and it has unit variance-covariance matrix. We will have the higher order moments of $U_{t,t+h}$ as functions of $(\mu,\sigma)$, if we assume that its distribution is elliptic. Alternatively, the conditional moments of asset returns can be assumed to be functions of some selected risk factors $f$. For example, $\mu$ and $\sigma$ can be computed via machine learning methods (see \cite{GKX2020}) or a Monte-Carlo simulation and clustering-based method introduced in \cite{Zhang2020b}. More detailed analysis of the choice of factors and a review of popular regression methodologies can be found in \cite{GKX2020}.

In empirical studies, we always have $|\alpha| = 1$, meaning that we will only compute the first order moments. The estimation of cross higher order moments, for example, the second order moments, also known as the covariance matrix, is not necessary. To understand this, consider a lead-lag panel regression of $\left(R_{t+1}^i\right)^2 = g\left(t, f_t^i\right) + \varepsilon_{t,t+1}^i$. The second order moment of the weighted assets $w \cdot R$ can be represented by $\mathbb{E}_t \left[(w \cdot R_{t+1})^2\right] = g\left(t, f_t^w\right)$, where $f_t^w$ is the associated factor value of the synthetic asset $w \cdot R$. This is an interpolation problem when $0<w<1$, on which the machine learning methods work well. The same applies to higher order cross moments.

\subsection{Asset Space Decomposition}
In order to apply the bottom-up approach to construct the optimal portfolios, we first use a top-down (hierarchical) clustering method to decompose the asset universe into stratified sub-spaces. Starting from the sub-spaces of the lowest level, we obtain the optimal portfolio weights based on the parameters input from Section \ref{sec:ip} and the methodology illustrated in Section \ref{sec:po}. By working from the lowest to highest level, we will obtain the optimal portfolio weights corresponding to each of the sub-spaces, and therefore each asset. To be specific, suppose that we have a $K$-vector of asset specific factors $\left\{f^k\right\}_{k = 1}^K$, and denote the realized values by $f_t^{k,j}$, where $j$ denotes the $j$th asset and it ranges in $[1, n]$. Time $t$ ranges in $[1, T]$. Therefore, there are $n \times T$ observations of the $K$-dimensional factor. Use hierarchical clustering approach on those observations and compute which cluster each related asset in the universe belongs to at each time $t$. A more straightforward way to create clusters is to consider the actual sector that each asset belongs to and categorize them by the related industries. In addition to the clustering approach, we can create various criteria based on the factor values to partition the asset universe into small buckets, with the assets in each bucket presenting some similar behaviors\footnote{For example, we can calculate scores based on a set of predetermined factors for each asset, rank and divide the asset space by the scores.}.

\subsection{Portfolio Optimization}
\label{sec:po}
\subsubsection{Static Portfolio Optimization}
\subsubsection*{The Algorithm}
\label{sec:ta}
Assume that we are going to maximize an objective function $G(w, R)$\footnote{For example, the objective function can be a reward minus a coherent risk measure on $R$}, where $w$ denotes the portfolio weights and $R$ is the rate of return vector of a certain class of assets. There are constraints on $w$, namely
\begin{align}
\label{eq:constr1}
F(w) &\geq 0 \\
\label{eq:constr2}
H(w) &=    0 \\
\label{eq:constr3}
w    &\in  (a, b).
\end{align}
$F$ and $H$ are nonlinear functions of $w$. A general formulation of the static portfolio optimization problem can be found in \cite{Homescu2014}. In this section, we outline a simulation and machine learning-based approach to obtain the optimized weights $w$. Assume that Equation \eqref{eq:constr2} can be rewritten as
\begin{align}
\label{eq:constr4}
w^n = h \left( w^1, w^2, \cdots, w^{n-1} \right).
\end{align}
The method works as following:
\begin{enumerate}
	\item Generate $M \times (n-1)$ uniform random numbers $\left\{w_m^j\right\}_{j = 1, m = 1}^{n-1, M}$, which satisfy $a^j < w_m^j < b^j$
	\item Compute $w_m^n = h \left( w_m^1, w_m^2, \cdots, w_m^{n-1} \right)$ for $m = 1, 2, \cdots, M$
	\item Find a subset of $\left\{w_m^j\right\}_{j = 1, m = 1}^{n-1, M}$ that satisfies Equation \eqref{eq:constr1} and denote it by $\mathcal{S}^F$
	\item Use hierarchical or regular clustering method to decompose $\mathcal{S}^F$ into $K$ disjoint clusters, denoted by $\left\{\mathcal{S}_k^F\right\}_{k = 1}^K$
	\item Denote the center of $\left\{\mathcal{S}_k^F\right\}_{k = 1}^K$ by $\left\{\bar{w}^k\right\}_{k = 1}^K$ and compute $k^* = \mathsf{argmax}_{1 \leq k \leq K} \left[ G\left( \bar{w}^k, R \right) \right]$
	\item Use hierarchical or regular clustering method to decompose $\mathcal{S}_{k^*}^F$ into $K$ disjoint clusters, denoted by $\left\{\mathcal{S}_k^{F, k^*}\right\}_{k = 1}^K$
	\item Repeat Step $4$ to Step $6$ until convergence.
\end{enumerate}
It would be interesting to discuss the utility loss introduced by this hierarchical construction. It can be easily understood that, at the last level of clustering, we perform optimization for each of the subsets and the final global weights are proportional to the weights in each final cluster. Of course, this methodology is sub-optimal compared to the global optimization on the whole asset universe. However, we have gains in terms of a faster computational speed, less resources requirement, less severe propagation of estimation errors, and the elimination of the potential corner solutions or local optimum. Moreover, the portfolio optimization is done in a bottom-up way, but the clustering is top-down, therefore our methodology enjoys the benefit of both approaches. Moreover, the reason to use clustering approach in Step $4$ above is to reduce the computation burden when $M$ is very large and the evaluations of objective function are time consuming.

\subsubsection*{Theoretical Convergence}
In this section, we discuss the global convergence of the proposed approach based on three critical assumptions below.
\begin{assumption}[Completeness]
\label{assu:1}
The random number generator $\Upsilon$ satisfies the following. Suppose that the size of random numbers generated is $N$, and the random numbers generated by $\Upsilon$ form a set $\beta_N$. Then we have the fact that $\cup_{N=1}^\infty \beta_N$ is always dense in the compact set $\mathcal{S}^F$.
\end{assumption}
Assumption \ref{assu:1} ensures that, as we sample more random realizations for the portfolio weight vectors, any point in $\mathcal{S}^F$ is reachable with the samples generated.
\begin{assumption}[Existence and Uniqueness]
\label{assu:2}
The constrained optimization problem $O(w)$ introduced in Section \ref{sec:ta} has a unique solution.
\end{assumption}
\begin{assumption}[Continuity]
\label{assu:3}
The optimization problem is \emph{continuous} with respect to $\beta_N$. This means that
\begin{align}
\lim_{K \rightarrow \infty} \mathsf{argsup}_{\cup_{N=1}^K \beta_N}O(w) &= \mathsf{argsup}_{\cup_{N=1}^\infty \beta_N}O(w)
\end{align}
where $O(w)$ is the original optimization problem with constraints.
\end{assumption}
Then, combining the above two assumptions, we have the theorem below as our main theoretical result.
\begin{theorem}[Global Convergence]
Under Assumptions \ref{assu:1}, \ref{assu:2} and \ref{assu:3}, our algorithm output is convergent to the unique optimal solution.
\end{theorem}

\subsubsection{Dynamic Portfolio Optimization}
In a dynamic portfolio optimization problem, we try to solve the Bellman equation
\begin{align}
\label{eq:bellman}
V^\pi(s) &= R(s) + \gamma \times \left[ \max_\pi \sum_{s' \in S} P_{s, \pi(s)} (s') V^\pi(s') \right]
\end{align}
where $R$ is the immediate reward function, $\pi:\mathcal{S} \rightarrow \mathcal{A}$ is a mapping from the state space $\mathcal{S}$ to the action space $\mathcal{A}$ and is called the policy function. $P_{s, \pi(s)} (s')$ denotes the probability transition matrix and $V^\pi(s)$ is the value function. Last, $\gamma \in (0, 1)$ is the discount factor process, which is often taken as a constant. The goal is to find an optimal policy function $\pi$ such that the value function is maximized. Of course, in general, $\pi$ is nonlinear in both time $t$ and state $s$\footnote{We will often assume that time variable $t$ is included in state vector $s$.}. However, it can be approximated locally in an open and sufficiently small region by its tangent space, which is represented by a linear equation. Further suppose that the state space $\mathcal{S}$ and action space $\mathcal{A}$ are compact sub-spaces of Euclidean space and we can generate uniform random numbers in $\mathcal{S}$ and $\mathcal{A}$. Decompose $U = \mathcal{S} \cup \mathcal{A}$ into small disjoint sub-spaces $\left\{ U_k \right\}_{k = 1}^K$, and we have $\pi(s)|_{U_k} \cong \delta_k^0 + \delta_k^1 \cdot s + \varepsilon_k$, where $\varepsilon_k$ is the approximation error term. The functional form of $\pi$ is solely determined by $\left(\delta_k^0,\delta_k^1\right)$ for each $k$. For the transition probability matrix $P_{s, \pi(s)} (s')$, one way to represent it is to assume a parametric model $s_{t+1} = f \left( s_{0:t}, a_{0:t}, e_{t, t+1} \right)$, where $s_{0:t} = \left( s_0, s_1, \cdots, s_t \right)$ and likewise for $a_{0:t}$. To solve the optimization problem in Equation \eqref{eq:bellman}, we generate $M$ independent copies of $\left(\delta_k^0,\delta_k^1\right)_{k = 1}^K$, therefore $M$ different functional forms of $\pi$, and for each copy, compute the value function via Monte-Carlo simulation based on the data generating process for $s_t$ and Equation \eqref{eq:bellman}, and use the method proposed in Section \ref{sec:ta} to determine the best choice of $\left(\delta_k^0,\delta_k^1\right)_{k = 1}^K$ among the $M$ independent samples. Last, the data generating process for $s_t$ can, alternatively, be replaced by a non-parametric inference directly using historical relationship.

\section{Empirical Studies}
\label{sec:es}
\subsection{Simulated Data}
\label{sec:sd}
Assume that there is an $m$-dimensional vector process $f_t$, whose data generating process (DGP) is
\begin{align}
\label{eq:genDGP}
f_{t+h} &= g(f_t, \vartheta_t) + e_{t,t+h}^f
\end{align}
where $\vartheta$ is another stochastic process\footnote{For example, the DGP can be an ARMA-GARCH process and $\vartheta$ is therefore the stochastic variance.} and $\mathbb{E}_t \left[ e_{t,t+h}^f \right] = 0$. The asset return vector is denoted by $r_t$, which is $n$-dimensional. We have approximately the following regression relationship
\begin{align}
\label{eq:regrel}
r_t = h(f_t) + e_t^r.
\end{align}
Here $e_t^r$ is considered as a small perturbation term, which might be originated from missing factors or measurement errors. Further observe that
\begin{align}
\label{eq:condexp}
\mathbb{E}_t \left[ r_{t+h} \right] &= h(t,h,f_t,\vartheta_t)+ u_{t,t+h}^{r,f}
\end{align}
where $h$ is potentially a nonlinear function of $(f,\vartheta)$ and $u_{t,t+h}^{r,f}$ is the pricing error term. The detailed configurations are described below. The DGP for the factor process of each stock is chosen as an ARMA-GARCH model
\begin{align}
\label{eq:ag}
f_t        &= \mu + \phi f_{t-h} + \sigma_t \epsilon_t \\
\sigma_t^2 &= \alpha + \beta \sigma_{t-h}^2 + \gamma f_{t-h}^2.
\end{align}
The factor $f$ is $n$-dimensional\footnote{This means the factor is asset specific and we have only $m = 1$ factor for each stock.}, $0<\mu<0.05$ is $n$-dimensional, $0<\phi<1$ is $n \times n$, $\sigma$ is $n \times 1$ and the error term $\epsilon_t = P \cdot u_t$, where the correlation generator $P$ is an $n \times n$ lower triangular matrix with squared sum of each row being $1$. $u_t$ is an $n$-dimensional independent Gaussian process with mean $0$ and variance $1$. The parameter set $(\mu,\phi,\alpha,\beta,\gamma,P)$ is generated randomly according to uniform distributions and the values ensure that the ARMA-GARCH models are stationary. The $n$-dimensional return process satisfies $r_t = 0.02 \times \sin(f_t) + \epsilon_t$\footnote{This functional form is to ensure that the return series generated are mostly around $(-0.02, 0.02)$.} and $\epsilon \cong \mathsf{Unif}(-0.0015, 0.0015)$ is an $n$-dimensional uniformly distributed random vector serving as the perturbation term, accounting for missing factors or measurement errors. The number of factors is $1$, the number of assets $n = 1000$ and the number of time periods $T = 250$. The number of clusters is $\left[\sqrt{1000}\right]$, i.e., the integer part of $\sqrt{1000}$. The portfolio weights are constrained within $[0, 1]$ and sum up to $1$. The objective function is the classical mean-variance quadratic one. The equity curve of the out-of-sample optimization results is displayed in Exhibit \ref{fig:sim} below.
\begin{figure}[htbp]\caption{Equity Curve for Simulation Study.}
	\centering
	\includegraphics[totalheight=6cm]{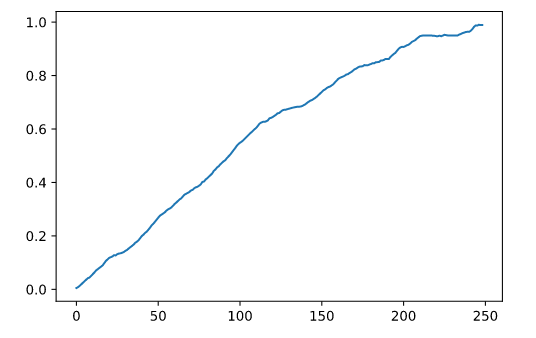}
	\label{fig:sim}
\end{figure}
It is obvious that the method performs well in an artificial simulation environment, according to the equity curve. The $x$-axis is the number of periods, which is up to $250$, and the $y$-axis is the value of the equity curve. From the plot we can see that the equity curve increase stably from $0$ to approximately $1$ with very limited drawdowns. This is inline with our expectations: in a simulation environment, we know and are able to recover exactly the functional relationship between factors and expected future returns. Of course, there are also small negative returns appearing along the equity curve. This is caused by the small perturbation term $e_t^r$ in Equation \eqref{eq:regrel} and the fact that the realization of future returns can deviate from their expected values, which is illustrated by the pricing error term $u_{t,t+h}^{r,f}$.

\subsection{U.S. Equity Market}
\label{sec:usdata}
\subsubsection{The Data}
The daily OHLC, trading volume and shares outstanding data are downloaded from WRDS for stocks traded in AMEX, NASDAQ and NYSE. The cross section contains $7596$ stocks. Time ranges from 20110103 to 20191231. The OHLC data are before dividend and stock splits. Therefore, we use the raw OHLC multiplied by the shares outstanding data to account for stock splits. In order for simplicity, we ignore the dividend effect. Because the portfolio weights are restricted between $0$ and $1$, the actual performance of the methodology should be better than what are presented.

\subsubsection{The Methodology}
To carry out the analysis, some details have to be determined. The objective function is $f(w) = \frac{\mu(w) + \frac{1}{2}s(w)}{\sigma(w) + \frac{1}{2}k(w)}$, where $(\mu,\sigma,s,k)$ are the conditional expected return, empirical volatility, skewness and kurtosis of the portfolio $w$. $(\sigma, s, k)$ are empirical values computed for every simulated portfolio weight vector $w$ using past $12$ months' asset return data. An alternative objective function is based on CRRA (constant relative risk aversion) utility function on terminal wealth $f(w) = \mathbb{E}_t \left[ \frac{1-(1+R_{t+h}(w))^\gamma}{1-\gamma} \right]$, where $R_{t+h}(w)$ is one-step ahead portfolio return associated with weight vector $w$. Portfolio weights are constrained within $(0, 1)$ and they sum up to $1$. The conditional expected returns are estimated via a $42$-factor\footnote{The details of the factors are available upon request.} lead-lag regression model implemented with Python function \textsf{XGBRegressor} provided by module \textsf{xgboost}. The regression is done in a rolling window manner, with time length $200$ weeks. The prediction is based on a $20$-day time frame and the factor values are sampled every $5$ business days. The clustering is done by the scores computed via equal weights on the factor values. The cross section is either the largest $500$ or $1,600$ companies in AMEX, NYSE and NASDAQ by market capitalization. The optimal portfolios are computed at the beginning of each period and are held until the end of the period.

\subsubsection{The Results}
\begin{figure}[htbp]\caption{Equity Curves for U.S. Stock Market.}
	\centering
	\includegraphics[width=\textwidth]{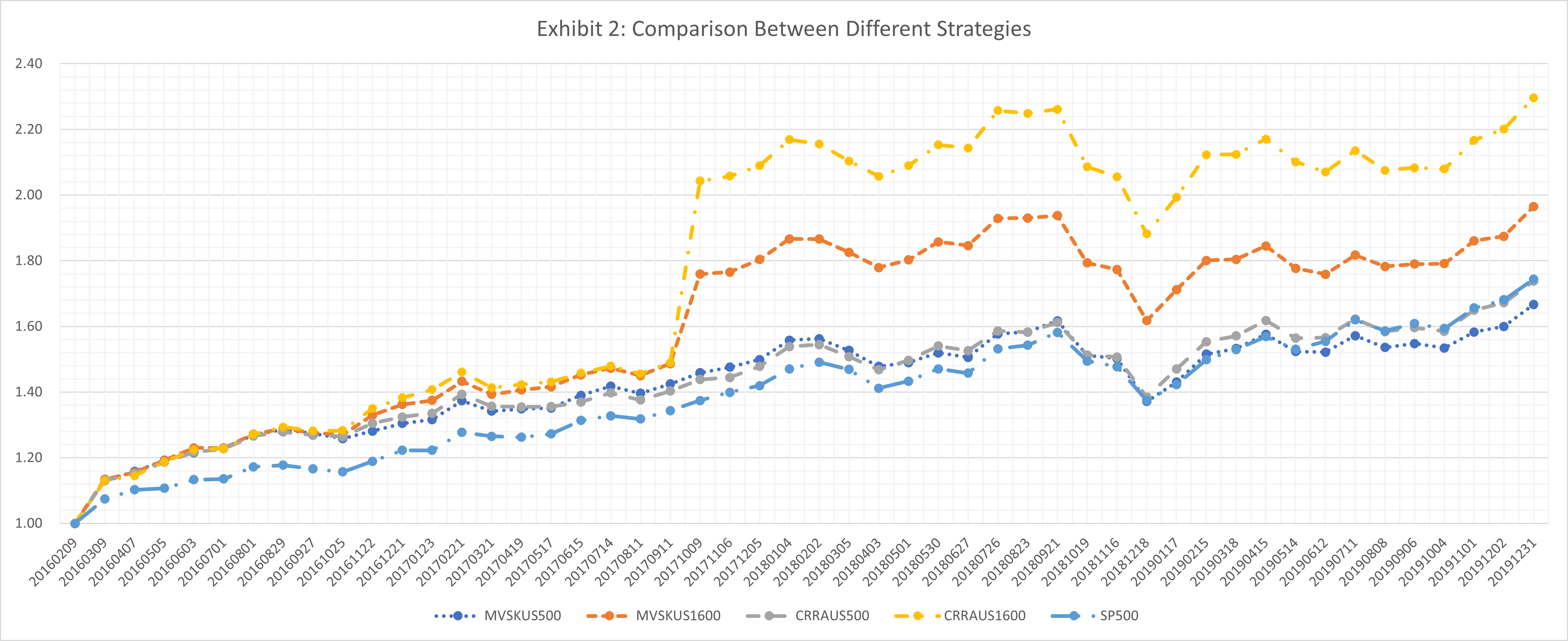}
	\label{fig:us}
\end{figure}

The backtesting results are summarized in the equity curve plot in Exhibit \ref{fig:us} and the performance metrics in Exhibit \ref{tab:us}. From Exhibit \ref{fig:us}, we can see that the CRRA1600 (Constant Relative Risk Aversion objective function optimization on the largest $1,600$ stocks in U.S. equity markets by market capitalization) performs best, with the terminal net value more than doubled compared to the initial capital. The second in place is MVSK1600 (Mean-Variance-Skewness-Kurtosis objective function optimization on the largest $1,600$ stocks in U.S. equity markets by market capitalization), with equity value approching $2.00$. The performance of MVSK500 and CRRA500 is close to S\&P500, with negligible excess returns. It can be seen from the plot that there is a jump in equity curve value at 20171009 for CRRA1600 and MVSK1600, causing the excess returns. In general, it can be concluded from our experiments that, in U.S. markets, performing the selected naive single period optimization schemes introduces little economic gains and excess returns compared to the market index. Exhibit \ref{tab:us} contains performance metrics, where \emph{Return} denotes annualized average arithmetic returns, \emph{Vol} denotes annualized standard deviation of the return series, \emph{IR} denotes information ratio, \emph{SR} represents Sortino ratio, \emph{CR} is Calmar ratio and \emph{MDD} is the abbreviate for max drawdown. It is surprising that none of the strategy information ratios exceed that of S\&P500. However, the annualized returns of some of the curves beat the financial market index.
\begin{table}\caption{Performance Metrics in U.S. Stock Market.}
	\small
	\centering
	\begin{tabular}{l *{7}{d{2.2}}}
		\hline
		\mc{Index} & \mc{Return} & \mc{Vol} & \mc{IR} & \mc{SR} & \mc{MDD}     & \mc{CR}    \\
		\hline\hline
		MVSK500    & 13.20\%     & 11.27\%  & 1.17    & 1.58    & 15.17\%      & 0.87       \\
		\hline
		MVSK1600   & 17.66\%     & 14.59\%  & 1.21    & 1.98    & 16.51\%      & 1.07       \\
		\hline
		CRRA500    & 14.23\%     & 11.28\%  & 1.26    & 1.78    & 14.10\%      & 1.01       \\
		\hline
		CRRA1600   & 22.49\%     & 21.49\%  & 1.05    & 2.60    & 16.78\%      & 1.34       \\
		\hline
		S\&P500    & 14.23\%     &  9.30\%  & 1.52    & 2.03    & 13.09\%      & 1.08       \\
		\hline
	\end{tabular}
	\label{tab:us}
\end{table}

\subsection{China A Share Market}
\label{sec:chndata}
\subsubsection{The Data}
The adjusted daily stock OHLC and trading volume data in CSI300 and CSI800 indexes are downloaded from Wind terminal. Time ranges from 20120206 to 20200928.

\subsubsection{The Methodology}
The objective function is $f(w) = \frac{\mu(w) + \frac{1}{2}s(w)}{\sigma(w) + \frac{1}{2}k(w)}$, where $(\mu,\sigma,s,k)$ are the conditional expected return, empirical volatility, skewness and kurtosis of the portfolio $w$. $(\sigma, s, k)$ are empirical values computed for every simulated portfolio weight vector $w$ using past $12$ months' asset return data. An alternative objective function is based on CRRA (constant relative risk aversion) utility function on terminal wealth $f(w) = \mathbb{E}_t \left[ \frac{1-(1+R_{t+h}(w))^\gamma}{1-\gamma} \right]$, where $R_{t+h}(w)$ is one-step ahead portfolio return associated with weight vector $w$. Portfolio weights are constrained within $(0, 1)$ and they sum up to $1$. The conditional expected returns are estimated via a $42$-factor\footnote{The details of the factors are available upon request.} lead-lag regression model implemented with Python function \textsf{XGBRegressor} provided by module \textsf{xgboost}. The regression is done in a rolling window manner, with time length $200$ weeks. The prediction is based on a $20$-day time frame and the factor values are sampled every $5$ business days. The clustering is done by the scores computed via equal weights on the factor values. The cross section is either the CSI300 or CSI800 index stocks. The optimal portfolios are computed at the beginning of each period and are held until the end of the period.

\subsubsection{The Results}
\begin{figure}[htbp]\caption{Equity Curves for China A Share Market.}
	\centering
	\includegraphics[width=\textwidth]{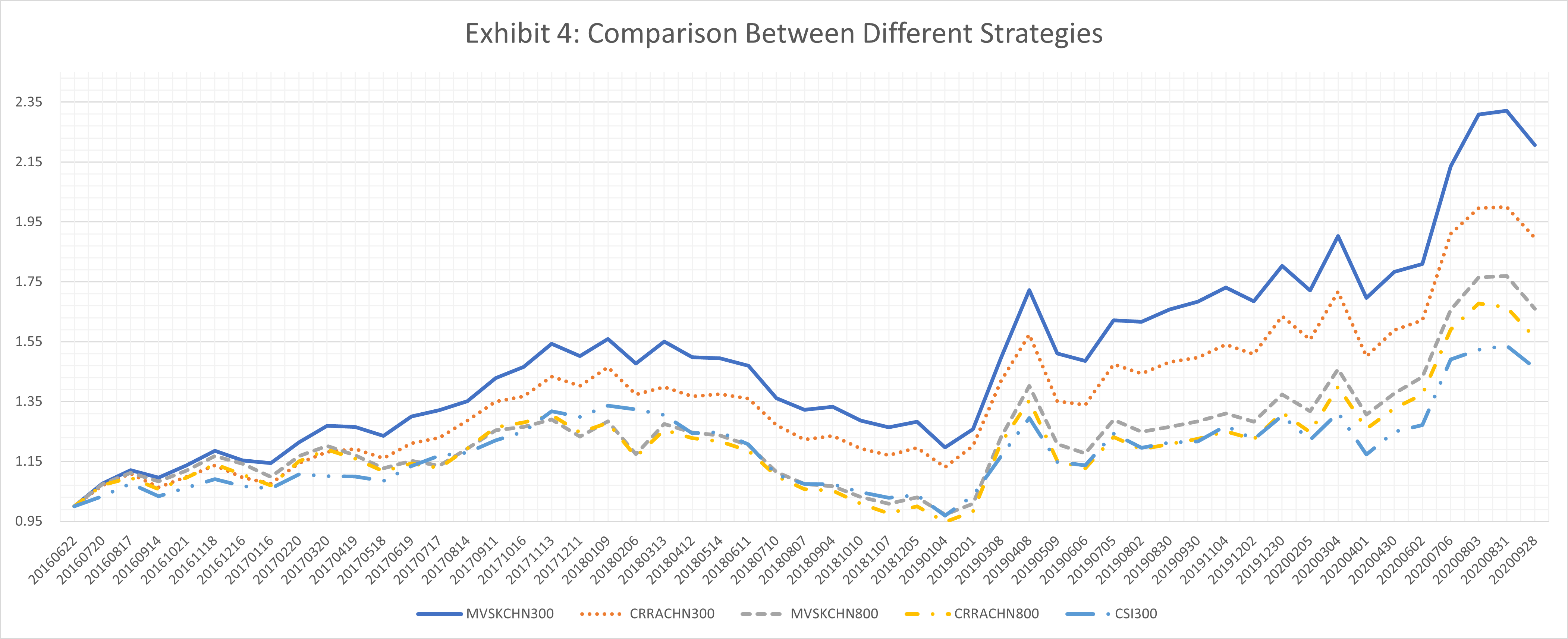}
	\label{fig:chn}
\end{figure}

\begin{figure}[htbp]\caption{Equity Curve of Excess Returns in China A Share Market.}
	\centering
	\includegraphics[width=\textwidth]{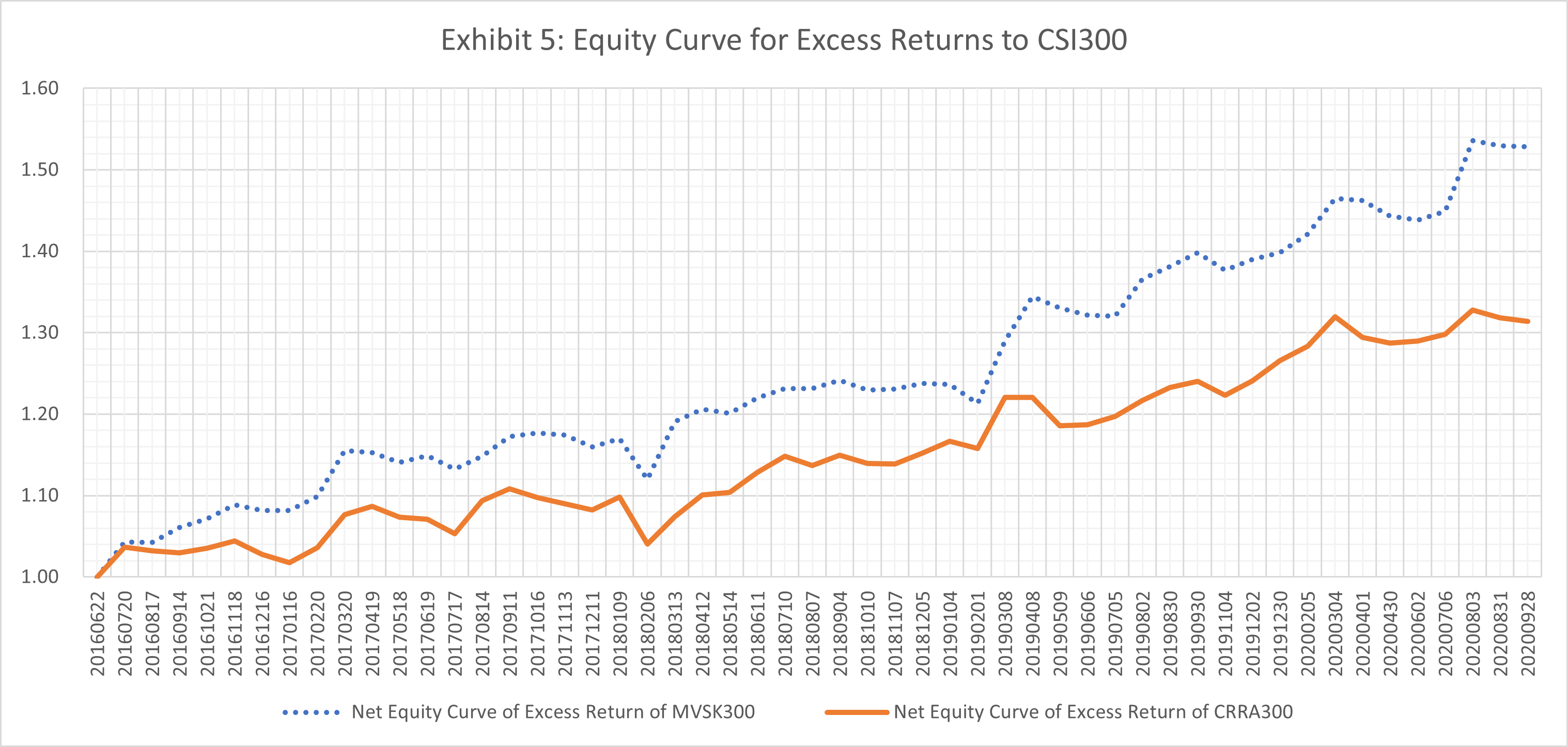}
	\label{fig:chnexcess}
\end{figure}

The backtesting results are summarized in the equity curve plots in Exhibits \ref{fig:chn} and \ref{fig:chnexcess} and the performance metrics in Exhibit \ref{tab:chn}. In China A shares market, the simple single period optimization methods reveal significant excess returns as can be observed in Exhibits \ref{fig:chn} and \ref{fig:chnexcess}. The MVSK300 scheme tops the horse-race, followed by the CRRA300. Although beaten by the aforementioned two schemes, MVSK800 and CRRA800 curves are still above the CSI300 index curve. Exhibit \ref{fig:chnexcess} documents significant and stable excess returns, which are positive through time. Exhibit \ref{tab:chn}, again, documents the performance metrics. Various risk and reward indexes point out that MVSK300 is the best strategy among the four competing methods and CSI300 bears minimum IR, SR and CR. 
\begin{table}[htbp]\caption{Performance Metrics in China A Share Market.}
	\small
	\centering
	\begin{tabular}{l *{7}{d{2.2}}}
		\hline
		\mc{Index} & \mc{Return} & \mc{Vol} & \mc{IR} & \mc{SR} & \mc{MDD}     & \mc{CR}    \\
		\hline\hline
		MVSK300    & 20.60\%     & 21.52\%  & 0.96    & 1.85    & 23.24\%      & 0.89       \\
		\hline
		MVSK800    & 14.41\%     & 22.49\%  & 0.64    & 1.30    & 24.62\%      & 0.59       \\
		\hline
		CRRA300    & 15.54\%     & 21.26\%  & 0.73    & 1.28    & 22.75\%      & 0.68       \\
		\hline
		CRRA800    & 13.05\%     & 22.95\%  & 0.51    & 1.15    & 27.40\%      & 0.48       \\
		\hline
		CSI300     & 10.51\%     & 18.87\%  & 0.56    & 0.97    & 27.53\%      & 0.38       \\
		\hline
	\end{tabular}
	\label{tab:chn}
\end{table}

\subsection{Stability of the Methodology}
\begin{figure}[htbp]\caption{Empirical Stability Plot.}
	\centering
	\includegraphics[width=\textwidth]{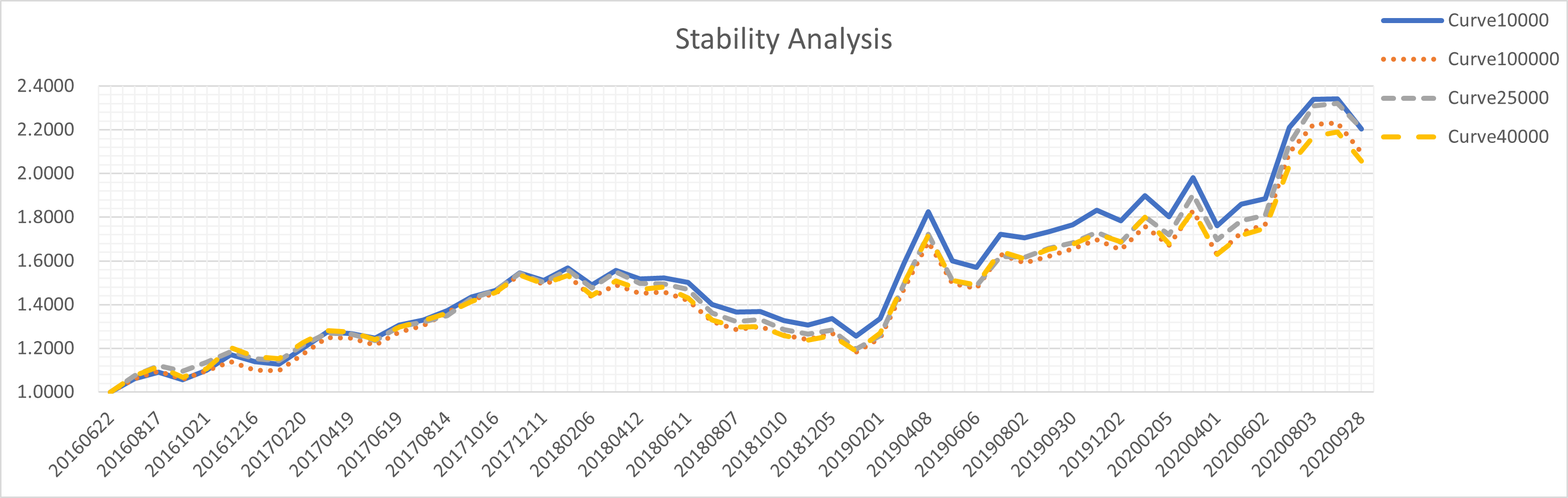}
	\label{fig:stability}
\end{figure}

Monte-Carlo simulation is used to construct the optimal portfolios. Therefore, a natural question to ask is whether the method is stable for different random numbers generated and how many samples are considered enough? In this section, we try to answer the question empirically only, although a theoretical derivation of the \emph{error bounds} for different samples $M$ is possible. In the sequel, we compare the empirical results on MVSK300 in China A share market, with $M$ ranges in $(10,000;25,000;40,000;100,000)$. The results with $M=100,000$ are taken as the benchmark values and RMSE's (root-mean-squared-error), as well as RMSRE's (root-mean-squared-relative-error), are computed with respect to the benchmark equity curve. We provide both the comparisons between different equity curves graphically, in Exhibit \ref{fig:stability}, and RMSE's/RMSRE's numerically in Exhibit \ref{tab:stability}. It can be observed that, with $40,000$ simulated weights, the result is close enough to that of $100,000$ samples. This can be considered as a \emph{convergence test} in the language of model validation.
\begin{table}[htbp]\caption{Empirical Stability Analysis.}
	\small
	\centering
	\begin{tabular}{l *{4}{d{2.2}}}
		\hline
		\mc{Index} & \mc{10,000} & \mc{25,000} & \mc{40,000} \\
		\hline\hline
		RMSE       & 0.08        & 0.04        & 0.03        \\
		\hline
		RMSRE      & 5.73\%      & 2.79\%      & 1.86\%      \\
		\hline
	\end{tabular}
	\label{tab:stability}
\end{table}

\section{Conclusion}
\label{sec:c}
In this paper, inspired by the methodology introduced in \cite{Zhang2020a}, we document a novel four-step portfolio optimization framework and test it with simulated and real financial data in China A-shares and U.S. equity markets. Our results reveal superior returns over the out-of-sample testing periods for both markets, which illustrates the usefulness of our methodology, that is not only a numerical framework, but also contributes to the literature of large scale optimizations. The empirical study of our proposed dynamic portfolio choice method via reinforcement learning is both interesting and important, yet postponed to future research. In addition, our methodology can be extended to fixed income and option portfolio selection, combining the work of \cite{Zhang2020b} and \cite{Zhang2020a}, which we leave to the interested readers as exercises.

\newpage
\bibliography{Reference}
\bibliographystyle{apalike}
\end{document}